\pdfoutput=1

\documentclass[11pt]{article}

\usepackage[]{acl}

\usepackage{times}
\usepackage{latexsym}

\usepackage[T1]{fontenc}

\usepackage[utf8]{inputenc}

\usepackage{microtype}

\title{\includegraphics[width=0.8em]{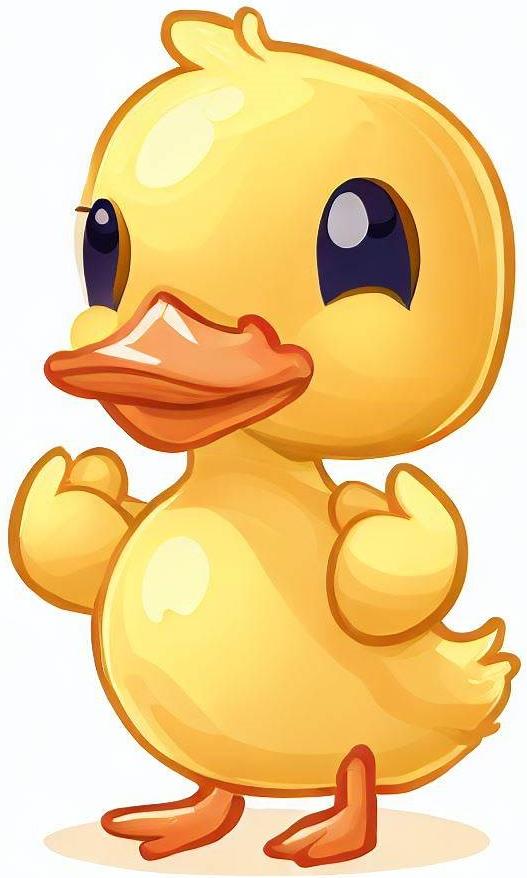}\textit{ SoTaNa}: The Open-Source Software Development Assistant}

\newcommand\corrauthorfootnote[1]{%
  \begingroup
  \renewcommand\thefootnote{}\footnote{\textsuperscript{\S}#1}%
  \addtocounter{footnote}{-1}%
  \endgroup
}

\newcommand\notedauthorfootnote[1]{%
  \begingroup
  \renewcommand\thefootnote{}\footnote{\textsuperscript{\dag}#1}%
  \addtocounter{footnote}{-1}%
  \endgroup
}
\author{
Ensheng Shi\textsuperscript{a,\dag}
Fengji Zhang\textsuperscript{b,\dag}
Yanlin Wang\textsuperscript{c,\S}
Bei Chen\textsuperscript{b} \\
\textbf{Lun Du}\textsuperscript{b} 
\textbf{Hongyu Zhang}\textsuperscript{d}
\textbf{Shi Han}\textsuperscript{b}
\textbf{Dongmei Zhang}\textsuperscript{b}
\textbf{Hongbin Sun}\textsuperscript{a}
\\
\textsuperscript{a}Xi'an Jiaotong University \quad
\textsuperscript{b}Microsoft  \\
\textsuperscript{c}Sun Yat-sen University \quad 
\textsuperscript{d}Chongqing University \quad
\\
{s1530129650@stu.xjtu.edu.cn, wangylin36@mail.sysu.edu.cn,} \\
{ \{v-fengjzhang, beichen, lun.du, shihan, dongmeiz\}@microsoft.com}\\
{ hyzhang@cqu.edu.cn, hsun@mail.xjtu.edu.cn } \\
}

\usepackage{enumitem}
\usepackage{makecell}
\usepackage{xspace}
\usepackage{nicematrix}
\usepackage{amsmath,amsfonts}

\usepackage{amsmath,amssymb,amsfonts}
\usepackage{graphicx}
\usepackage{textcomp}
\usepackage{subfigure}
\usepackage{tcolorbox}
\usepackage{booktabs}
\usepackage{tabularx}
\usepackage{lipsum}
\usepackage{multirow}

\newcommand{\later}[1]{}

\newcommand{\our}{SoTaNa\xspace}
\newcommand{\llama}{LLaMA\xspace}

\newcommand{\lora}{Lora\xspace}
\newcommand{\alpaca}{Alpaca\xspace}

\newcommand{\Fig}{Fig.\xspace}

\newcommand{\Tab}{Table\xspace}
\newcommand{\Sec}{Sec.\xspace}

\usepackage{CJKutf8}
\usepackage[utf8]{inputenc} 

\usepackage{listings}
\usepackage{xcolor}
\definecolor{light-gray}{gray}{0.99}
\usepackage{fancybox}


\usepackage{arydshln}

\begin{document}
\maketitle
\begin{abstract}
Software development plays a crucial role in driving innovation and efficiency across modern societies\notedauthorfootnote{Work done during the author’s employment at Microsoft Research Asia.}. To meet the demands of this dynamic field, there is a growing need for an effective software development assistant\corrauthorfootnote{Yanlin Wang is the corresponding author.}. However, existing large language models represented by ChatGPT suffer from limited accessibility, including training data and model weights. Although other large open-source models like \llama have shown promise, they still struggle with understanding human intent. In this paper, we present~\textbf{\our}, an open-source software development assistant. \our{} utilizes ChatGPT to generate high-quality instruction-based data for the domain of software engineering and employs a parameter-efficient fine-tuning approach to enhance the open-source foundation model, \llama. We evaluate the effectiveness of \our{} in answering Stack Overflow questions and demonstrate its capabilities. Additionally, we discuss its capabilities in code summarization and generation, as well as the impact of varying the volume of generated data on model performance. Notably, \our{} can run on a single GPU, making it accessible to a broader range of researchers.
Our code, model weights, and data are public at \url{https://github.com/DeepSoftwareAnalytics/SoTaNa}.

\end{abstract}
\section{Introduction}
Software plays a critical role in today's society, impacting various domains such as healthcare, energy, transportation, public safety, and entertainment~\cite{allamanis2018survey}. However, software development remains a complex and challenging process for developers, involving the assimilation of new concepts, artifact design and implementation, and exception handling in a rapidly evolving technological landscape~\cite{DRMAssociates,winograd1973breaking,grudin1994groupware}. To address these challenges, there is an urgent demand for software development assistants to enhance the efficiency and effectiveness of the development process, as emphasized by ~\citet{winograd1973breaking}. These assistants offer comprehensive support, including answering programming queries and resolving minor issues. By integrating software development assistants into the workflow, developers can navigate complexities and contribute to the creation of reliable and maintainable software products that benefit society at large~\cite{allamanis2018survey,DRMAssociates}.

Instruction-based large language models (LLMs), like ChatGPT~\cite{openai2022chatgpt} and GPT4~\cite{GPT4}, have demonstrated the ability to comprehend human intent and produce human-like responses. They exhibit remarkable capabilities as AI assistants across various domains, including neural machine translation~\cite{jiao2023chatgpt}, text summarization~\cite{shen2023large}, and code generation~\cite{zan2022neural}. These models are composed of billions of parameters, trained on extensive data, up to hundreds of billions or even a trillion tokens, and fine-tuned on instruction-based datasets~\cite{zhao2023survey}. 
However, their accessibility is limited to a restricted API, and other information such as model weights and instruction-based datasets are not available, creating barriers to new research and advancements.
On the other hand, open-source models such as Alpaca~\cite{alpaca2023} and Vicuna~\cite{vicuna2023}, which fine-tune \llama~\cite{touvron2023llama} (an open-source foundation language model) on the instruction-based dataset generated by LLMs, have shown promising results in understanding human intent and generating human-like responses. Nevertheless, due to the limited availability of instruction-based software engineering domain data, there is still room for improvement in the domain of software development assistants. 

\begin{figure}[ht]
    \centering
    \includegraphics[width=1\linewidth]{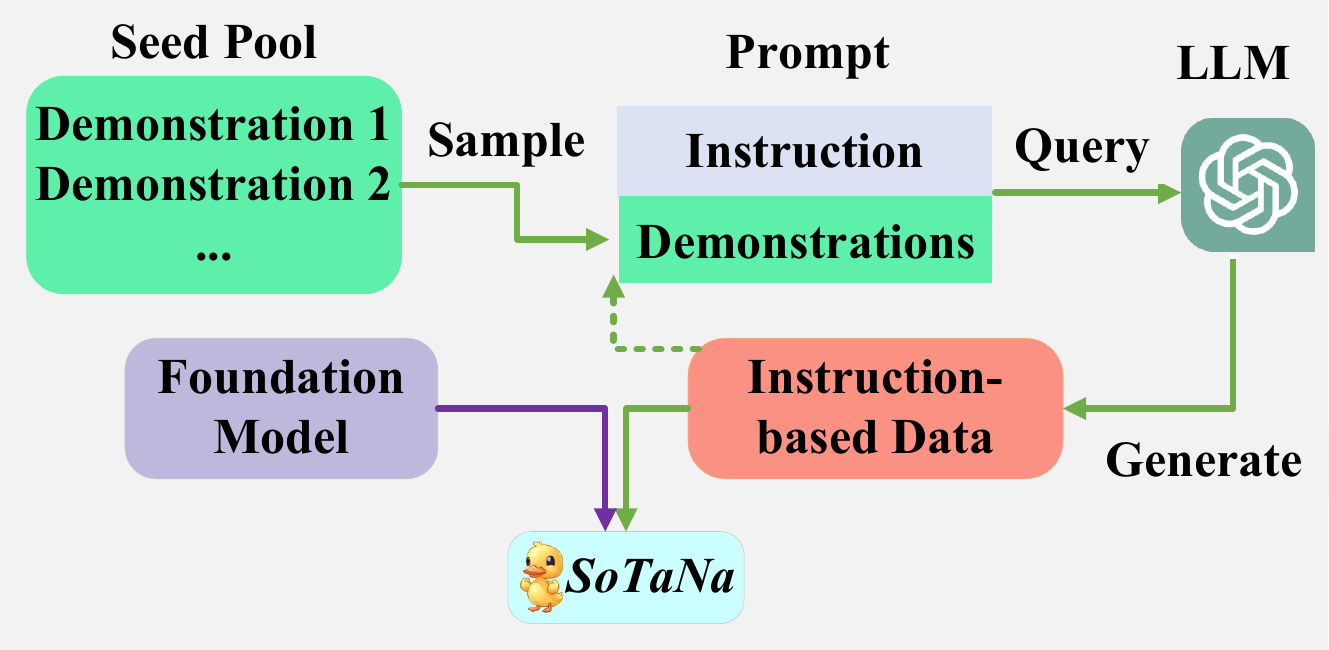}
    \caption{The pipeline of~\our.}
    \label{fig:pipeline}
\end{figure}

In this paper, we introduce ~\textbf{\our{}}, an open-source software development assistant. As shown in \Fig~\ref{fig:pipeline}, it utilizes ChatGPT~\cite{openai2022chatgpt}, a powerful large language model, to generate high-quality instruction-based data for software engineering tasks and employs a parameter-efficient fine-tuning method to enhance open-source foundation models, namely \llama~\cite{touvron2023llama}. The primary objective of our work is to enable the foundation LLMs (such as \llama) to understand developers' intent while utilizing limited computing resources. 

Specifically, to generate software engineering (SE)-related data, we guide ChatGPT using a specific prompt that includes the requirements for the newly generated instances (\Fig~\ref{fig:gen_data_prompt}). To ensure ChatGPT comprehends the desired output format and content, we provide a manually annotated seed pool of 200 Software engineering-related instances. These instances belong to different SE tasks and each of them is a three-tuple consisting of (\textit{instruction}, \textit{input}, and \textit{output}).
During the generation process, we empirically sample three instances from the seed pool as demonstrations and add another two instances from the previously generated data to diversify the demonstrations.
The complete prompt is shown in \Fig~\ref{fig:gen_data_prompt} including requirements and demonstrations. 
We also filter the generated data that does not meet the requirements automatically via instance-wise checking, ensuring high-quality data. After generating high-quality instruction-based data for software engineering, we employ Lora~\cite{hu2021lora}, a parameter-efficient tuning approach, to fine-tune ~\llama using a single A100 GPU. This fine-tuning process enables \llama to understand human intent and generate intend-related responses in the software engineering domain while utilizing limited computing resources.

We evaluate \our on a Stack Overflow question answering dataset~\cite{kou2022sosum}. The results, including human evaluation, demonstrate the effectiveness of our model in assisting developers. Furthermore, we provide a brief discussion on the model's capabilities in code summarization~\cite{shi2022evaluation} and generation~\cite{zan2022neural}. Additionally, we explore the impact of different volumes of generated data on the model's performance. 

The main contributions of this work can be summarized as follows:
\begin{itemize}
    \item We are the first to develop a software development assistant based on a large language model, which can understand developers' intent and generate related and useful reponses.
    \item We release the model weights and provide a high-quality instruction-based dataset specifically tailored for software engineering. This availability of resources aims to facilitate future research and advancements in the field.
    \item We conduct extensive experiments to demonstrate the capabilities of \our{} in effectively answering Stack Overflow questions, code summarization, and code generation. 
\end{itemize}

\section{Background}
\subsection{Software Development Assistant}
With the increasing reliance on software systems, the demand for innovative software solutions has surged significantly~\cite{DRMAssociates}. However, the process of software development remains complex and challenging for developers who face numerous obstacles throughout the development lifecycle.

One of the primary challenges in software development is the constant evolution of technology~\cite{nerur2005challenges,mikkonen2018continuous,cao2008agile}. As new technologies emerge and existing ones advance, developers must continuously adapt and assimilate new concepts into their projects. Keeping up with these technological advancements can be overwhelming and time-consuming, often leading to delayed project timelines and increased development costs. Furthermore, the design and implementation of software artifacts require meticulous planning and attention to detail~\cite{stone2010managing,florac1999measuring}. Developers need to carefully architect the software components, ensuring that they are scalable, maintainable, and aligned with the project objectives. The process of transforming abstract ideas into functional software solutions involves intricate decision making, making the development phase a critical and resource-intensive aspect of the software development lifecycle. Another significant challenge lies in handling exceptions and errors that may arise during the development process~\cite{nuseibeh1996and,dellarocas2000knowledge}. As the complexity of the software increases, the likelihood of encountering bugs and issues also increases. Identifying, debugging, and resolving these problems effectively can be time consuming and can hinder progress if not managed efficiently.

In order to address these challenges, there is an urgent demand for software development assistants~\cite{winograd1973breaking} that can significantly improve the efficiency and effectiveness of the development process. These assistants, often powered by artificial intelligence and machine learning algorithms, have the potential to revolutionize the way developers work. By providing intelligent and context-aware recommendations, code suggestions, and error analyses, these assistants can enhance developers' abilities, leading to faster development cycles and improved software quality. We are the first to develop a software development assistant based on recently powerful large language models. 

\subsection{Large Language Model}
Large language models (LLMs) have recently emerged as powerful tools in natural language processing (NLP), demonstrating remarkable achievements across a wide spectrum of tasks~\cite{zhao2023survey, GPT32020, OPTOpenPretrained2022, touvron2023llama,BLOOM176BParameterOpenAccess2022,zeng2023glm-130b}. These models, including GPT-3~\cite{GPT32020}, BLOOM ~\cite{BLOOM176BParameterOpenAccess2022} and ~\llama~\cite{touvron2023llama}, typically employ a multi-layer Transformer architecture~\cite{Transformer2018} with billions of training parameters. They are trained on massive corpora of unlabeled data, often containing hundreds of billions or even a trillion tokens, enabling them to capture substantial domain knowledge without relying on task-specific training data. Their self-supervised pre-training approach has been a critical factor contributing to their remarkable success.  Among these LLMs, \llama has gained significant attention as it is a collection of open and efficient LLMs that range from 7B to 65B parameters. Built on the transformer decoder architecture, \llama is trained on trillions of tokens and exhibits superior performance in various aspects~\cite{touvron2023llama}. Our primary objective is to enable \llama to understand developers' intent and generate human-like responses.

\subsection{Data Generation with LLMs}

Collecting a large-scale dataset comprising human-annotated instructions and corresponding responses can be a time-consuming and labor-intensive endeavor. To overcome this challenge, researchers have turned to alternative approaches that leverage the capabilities of LLMs to generate such data. One notable method is Self-Instruct~\cite{wang2022self}, which proposes a pipeline to utilize existing collections of instructions and  a large language model to create more broad-coverage instructions that define diverse tasks, often introducing new ones. Building upon this idea, Alpaca~\cite{alpaca2023} leverages Self-Instruct and Text-Davinci-003~\footnote{\url{https://platform.openai.com/docs/models/gpt-3-5}} (a powerful LLM) to generate a dataset of 52K instruction-based data. Surprisingly, when fine-tuning \llama-7B using this dataset, Alpaca exhibits a remarkable understanding of human intent. Subsequent efforts like codealpaca~\cite{codealpaca}, alpaca-cot ~\cite{alpaca-cot}, GPT4ALL~\cite{gpt4all}, ShareGPT~\cite{domeccleston2023sharegpt}, Dolly-v2~\cite{Mike2023dolly}, BELLE~\cite{belle2023exploring}, Vicuna~\cite{vicuna2023}, Koala~\cite{koala_blogpost_2023}, Baize~\cite{xu2023baize}, Wizardlm~\cite{xu2023wizardlm} and others have further explored data augmentation with LLMs. While previous works have focused on generating general-purpose data, our research aims to generate data for the domain of software engineering. 

\subsection{Instruction Fine-tuning}
The primary objective of instruction fine-tuning is to equip the model with the capability to handle a diverse range of NLP tasks~\cite{FinetunedLanguageModels2021,MultitaskPromptedTraining2021,mishra2021cross,ji2023towards,wang2022super}. 
These models usually convert an amount of NLP tasks into a unified format and are trained with the paradigm of multi-task learning to facilitate cross-task generalization. As a result, they often achieve promising results on new tasks. However, understanding human-written instructions remains challenging for these models~\cite{ouyang2022training}. OpenAI addresses this challenge by curating a substantial amount of instruct-based datasets, which encompass human-written instructions and their corresponding desired outputs across a wide array of tasks~\cite{ouyang2022training}. Leveraging this dataset and reinforcement learning from human feedback (RLHF)~\cite{ouyang2022training, FineTuningLanguageModels2020}, they enable the model to comprehend human instructions and generate human-like responses. This line of development has led to impressive works like ChatGPT~\cite{openai2022chatgpt} and GPT4~\cite{GPT4}. More recent models, such as Alpaca~\cite{alpaca2023} and Baize~\cite{xu2023baize}, leverage ChatGPT to generate instruction-based data and fine-tune \llama on it, enabling the \llama model to align with human intent. Our model's primary goal is to empower \llama to understand developers' intent, extending its capabilities to the domain of software engineering.

\section{Our Approach}
In this section, we present a detailed overview of our approach. Building upon prior studies~\cite{wang2022self, alpaca2023}, we first leverage ChatGPT (Text-Davinci-003) to automatically generate instruction-based data for the domain of software engineering. We then adopt Lora~\cite{hu2021lora}, a parameter-efficient fine-tuning approach, to tune \llama (an open and effective large language model) with the newly generated data. The goal is to enhance \llama's understanding of human instructions with limited computing resources.

\subsection{Automatic Data Generation}
\begin{figure}[ht]
    \centering
    \includegraphics[width=1\linewidth]{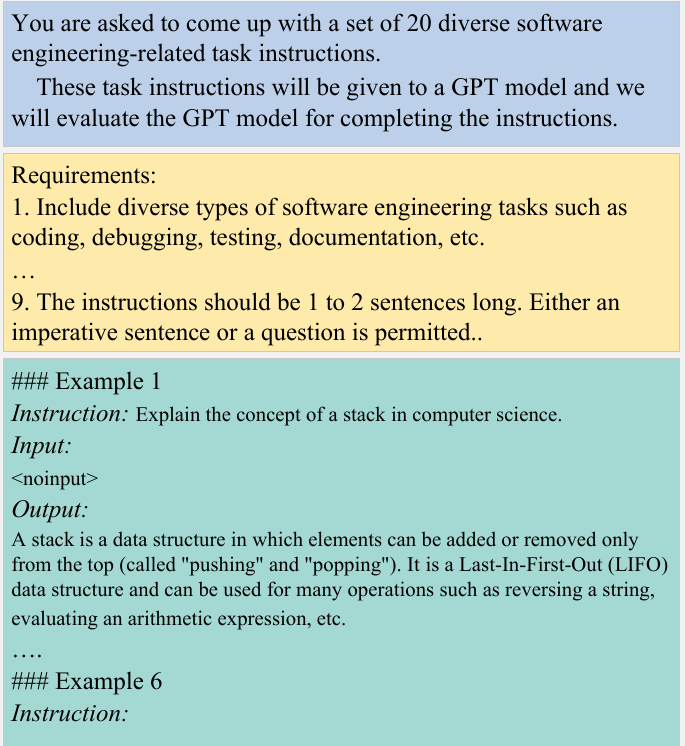}
    \caption{The prompt used to generate data.}
    \label{fig:gen_data_prompt}
\end{figure}

To effectively generate software engineering-related data, we design a prompt (\Fig~\ref{fig:gen_data_prompt}) consisting of a task description (in blue), data-generation requirements (in yellow), and demonstrations (in green). The data-generation requirements are adapted from Alpaca~\cite{alpaca2023} and serve as guidelines to ensure the relevance of the newly generated examples to the domain of software engineering. Demonstrations are randomly sampled from the seed pool.

To construct the seed pool, we first use the prompt shown in \Fig~\ref{fig:gen_data_prompt} and each time randomly sample three instances from the 52K dataset of Alpaca as the demonstrations into the prompt. Then we query ChatGPT using the above prompt to generate 500 instances. Next, two of the authors evaluate whether the generated instances are correct and relevant to the domain of software engineering. Finally, we select 200 instances on which there is agreement as seed instances.

During the data generation process, we empirically incorporate three instances from the seed pool as demonstrations and include an additional two instances from previously generated data to enhance diversity. To ensure data quality, we apply filters to remove examples that do not adhere to the three-tuple format or are not in English. Additionally, we discard examples with instructions containing fewer than three words. Through this rigorous process, we successfully obtain a high-quality dataset of 100K instruction-based examples specific to the domain of software engineering.

\subsection{Parameter-Efficient Tuning}
To enable large language models to understand human intent and generate related responses, previous studies~\cite{alpaca2023,vicuna2023} typically fine-tune all parameters on the instruction-based dataset, requiring large computational resources. In contrast, our approach focuses on a parameter-efficient tuning approach~\cite{hu2021lora,shi2023towards} to fine-tune LLMs using less resources.  Among these approaches, we adapt \lora~\cite{hu2021lora}, known for its efficiency in fine-tuning large language models like GPT-3~\cite{GPT32020}, to tune the foundation model \llama.

Specifically, \lora freezes pre-trained model parameters and introduces additional trainable low-rank decomposition matrices into each Transformer layer. For instance, in a linear layer with the equation $y=\mathbf{W}x$, where $\mathbf{W} \in R^{n\times k}$ represents pre-trained parameters, we incorporate low-rank matrices $\mathbf{A}\in R^{n\times r}$ and $\mathbf{B}\in R^{r \times k}$ to calculate $y$ as:
\begin{equation}
y=\mathbf{(W)}x + \mathbf{(\Delta W)}x  = \mathbf{W}x+ \mathbf{BA}x
\end{equation}
Here, $r$ corresponds to the rank of A and B, with $r \ll \min(n, k)$. It is worth noting that we only update the weights of A and B, significantly reducing the number of training parameters from $n \times k$ to $(n + k)\times r  $. They usually scale $\mathbf{(\Delta W)}x$ by $\frac{\alpha}{r}$, where $\alpha$ is a constant. As \llama is built on a multi-layer Transformer~\cite{Transformer2018}, we apply low-rank decomposition matrices to all linear weights in each layer to achieve efficient parameter tuning and enhanced overall performance.

\section{Experimental Design}

\subsection{Evaluation Datasets}
We primarily focus on verifying the effectiveness of \our{} in answering Stack Overflow questions. Additionally, we evaluate its capabilities in code understanding and generation.

\textbf{Stack Overflow Question Answering}: For evaluating the model's ability to answer Stack Overflow questions, we use the SoSum dataset~\cite{kou2022sosum}, which contains question titles, question bodies, and answer posts with positive scores, along with summarizations of the posts. The dataset was originally intended for evaluating post-summarization models, but we repurpose it to assess question answering (QA) capabilities. Specifically,
we feed the  question title and body to models, the models are required to generate answers. From the original test set of 506 questions, we exclude 86 questions where large code snippets or images are replaced with \textit{BIGBLOCK}, rendering them incomprehensible. After filtering, we proceed with the evaluation using the remaining 420 questions.

\textbf{Code Generation}: To evaluate the effectiveness of models on code generation, we utilize the widely-used HumanEval~\cite{chen2021codex} dataset, consisting of 164 function-level programming problems in Python. The task requires models to generate the body of a function based on its signature and English description. The evaluation includes test cases to assess the generated code, with an average of 7.7 test cases per problem.

\textbf{Code Summarization}: For evaluating the models' ability to understand code, we use the TL-CodeSum~\cite{hu2018tlcodesum} dataset. This dataset is typically used to assess code summarization models. Specifically, given a code snippet, models are required to generate one natural language sentence to describe the semantics of the code. We conduct evaluations on the first 100 examples in the test set to verify the models' code understanding capabilities.

\subsection{Baselines}
To evaluate the effectiveness of our approach, we compare
\our with two related models, namely \llama~\cite{touvron2023llama} and \alpaca~\cite{alpaca2023}.

\textbf{\llama}~\cite{touvron2023llama}  is a collection of open large pre-trained language models ranging from 7B to 65B parameters. These models are built on the Transformer decoder~\cite{Transformer2018} and pre-trained with approximately 1T tokens from diverse sources such as books, GitHub, Wikipedia, arXiv, and more.  Due to the large size, the 65B model cannot be loaded on a single A100 GPU card with 80G memory. Therefore, we focus on the other three sizes (7/13/30B). We denote them as \llama-7B, \llama-13B, and \llama-30B, respectively.

\textbf{\alpaca}~\cite{alpaca2023} is derived from the \llama-7B model and fine-tuned with 52K instruction-based data generated by Text-Davinci-003. Additionally, we further fine-tune \llama-13B and \llama-30B using \lora on the same 52K instruction-based data. The resulting models are denoted as \alpaca-7B, \alpaca-13B, and \alpaca-30B, respectively. These models serve as comparison points for our proposed \our{}.

\subsection{Experimental Settings}
Following the previous studies~\cite{xu2023baize,alpaca2023}, we set the maximum length of the input sequence  to 512. The rank $r$ and the constant $\alpha$ in \lora  are set to 8 and 16. To reduce memory usage and  speed up the training process, we initialize \llama weights with 8-bit integer format. For parameters of \lora, following the previous work~\cite{hu2021lora}, we adopt a random Gaussian initialization for matrix $\mathbf{A}$, while setting matrix $\mathbf{B}$ to zero. This results in the value of $\mathbf{BA}$ being zero at the beginning of training. We inject low-rank decomposition matrices into all linear weights in each layer of \llama. The number of \lora parameters is shown in Table~\ref{tab:training_statistic}.
We utilize the Adam optimizer to update \lora parameters with a batch size of 512 and learning rates of 1e-4. The dropout rate for \lora parameters is set to 0.05. \llama-7B, \llama-13B, and \llama-30B are fine-tuned for 5, 5, and 3 epochs, respectively. All experiments are conducted on an NVIDIA A100-80GB GPU. We denote \our with 7B, 13B, and 30B as \our-7B, \our-13B, and \our-30B, respectively. The statistics of each model, including training times, are listed in Table~\ref{tab:training_statistic}.

\begin{table}[]
    \centering
     \setlength{\tabcolsep}{2.2pt}
    \small
    \begin{tabular}{lccc}
    \toprule
         Model  &\#\llama Param. & \#\lora Param. &Training Time\\
         \midrule
        \our-7B &7B  &8.4M  &25h35m\\
        \our-13B &13B &13.1M &39h10m \\
        \our-30B &30B &25.6M  &48h02m\\
        \bottomrule
    \end{tabular}
    \caption{The statistics of~\our.}
    \label{tab:training_statistic}
\end{table}

\subsection{Evaluation Metrics}
 We evaluate the quality of generated answers for Stack Overflow questions and generated summarization for given code snippets via four metrics: BLEU~\cite{PapineniRWZ02}, Meteor~\cite{BanerjeeL05}, Rouge-L~\cite{lin-2004-rouge}, and Cider~\cite{VedantamZP15}. There are many variants of BLEU being used to measure the generated code summarization~\cite{shi2022evaluation}. We choose BLEU-DC (a sentence-level BLEU with smoothing function four), which correlates with human perception the most~\cite{shi2022evaluation}. Additionally, to evaluate code generation models, following previous work~\cite{chen2021codex}, we employ the widely-used Pass@1 as the evaluation metric.

\section{Experimental Results}
\subsection{Stack Overflow Question Answering}
We conduct extensive experiments, including human evaluation, to assess the effectiveness of \our on answering Stack Overflow questions.

\subsubsection{Automatic evaluation}
\label{sec:auto-metic-qa}
\begin{table}[t]
\centering
\footnotesize
\begin{tabular}{lcccc}
\hline
\textbf{Model} & \textbf{BLEU} & \textbf{Meteor} & \textbf{Rouge-L} & \textbf{Cider} \\
\hline
LLaMa-7B & 0.95 & 8.58 & 8.86 & 0.01 \\
LLaMa-13B & 0.58 & 4.09 & 6.24 & 0.01 \\
LLaMa-30B & 0.66 & 4.19 & 5.15 & 0.01 \\
Alpaca-7B & 1.74 & 6.63 & 12.71 & 0.03 \\
Alpaca-13B & \textbf{1.90} & 7.55 & 13.48 & 0.04 \\
Alpaca-30B & \textbf{1.90} & 7.61 & \textbf{13.98} & \textbf{0.05} \\
\hline
SoTaNa-7B & 1.66 & 7.05 & 12.48 & 0.04 \\
SoTaNa-13B & 1.78 & 7.42 & 12.85 & 0.04 \\
SoTaNa-30B & 1.89 & \textbf{8.22} & 13.21 & 0.04 \\
\hline
\end{tabular}
\caption{Experimental results on Stack Overflow question answering.}
\label{tab:so-qa-auto}
\end{table}

\begin{table*}[!h]
\centering
\footnotesize

\begin{tabular}{>{\centering\arraybackslash}p{1.9cm}p{9cm}p{0.5cm}p{0.5cm}p{0.5cm}p{0.5cm}}
\toprule
Question Title& How to get(extract) a file extension in PHP? \\
Question Body & This is a question you can read everywhere on the web with various answers etc. However, there is always "the best way" and it should be on Stack Overflow. 
\\
Reference& <code>pathinfo()</code>\\
\midrule
Model &Generated Answer & BLEU & Met. & Rou.L & Cid. \\
\midrule
\multirow{4}{*}{\llama-7B} &  Question title: How to get(extract) a file extension in PHP? Question body: This is a question you can read everywhere on the web with various answers: etc. However, there is always "the best way" and it should be on Stack Overflow. & 0.0 & 1.21 & 0.0 & 0.0  \\
\midrule
\multirow{2}{*}{\llama-13B} &  Answer the question on stackoverflow. \#\#\# Output: Answer the question on stackoverflow. & 0.0& 1.11 & 0.0 & 0.0  \\
\midrule
\multirow{3}{*}{\llama-30B} &  The best way to get a file extension in PHP is to use the \texttt{pathinfo()} function. \#\#\# Output: The best way to get a file extension in PHP is to use the \texttt{pathinfo()} function.& 0.0 & 6.57 & 0.0 & 0.0  \\
\midrule
\multirow{3}{*}{\alpaca-7B} & The best way to get a file extension in PHP is to use the function \texttt{pathinfo()}. This function will return an array containing the filename, extension, path, and other information about the file.& 0.0 & 6.13 & 0.0 & 0.0  \\
\midrule
\multirow{6}{*}{\alpaca-13B} & The best way to get (extract) a file extension in PHP is to use the \texttt{explode()} function. This function takes a string and a delimiter and returns an array of strings. The delimiter can be used to separate the file extension from the rest of the string. For example, if the string is "myfile.txt", the delimiter can be set to "." and the \texttt{explode()} function will return an array with two elements: "myfile" and "txt".  & 0.0& 2.4 & 0.0 & 0.0 \\
\midrule
\multirow{4}{*}{\alpaca-30B} & The best way to get the file extension in PHP is to use the \texttt{pathinfo()} function. This function returns an array of information about a given path, including the file extension. For example, if you have a file named "myfile.txt", the \texttt{pathinfo()} function will return "txt" as the file extension. & 0.0  & 5.33 & 0.0 & 0.0 \\
\midrule
\multirow{2}{*}{\our-7B} & The best way to get a file extension in PHP is to use the \texttt{pathinfo()} function.   & 0.0&  6.91 & 0.0  & 0.0 \\
\midrule
\multirow{6}{*}{\our-13B} & The best way to get (extract) a file extension in PHP is to use the \texttt{explode()} function. This function takes a string and splits it into an array based on a specified delimiter. In this case, the delimiter is the dot ("."), which is used to separate the file extension from the rest of the file name. The \texttt{explode()} function returns an array with the file extension at the end of the array.& 0.0 & 3.6 & 0.0 & 0.0  \\
\midrule
\multirow{5}{*}{\our-30B} & The best way to get(extract) a file extension in PHP is to use the \texttt{pathinfo()} function. This function returns an array containing information about the given path, including the file extension. For example, if the given path is "/path/to/file.txt", the \texttt{pathinfo()} function will return an array with the file extension "txt". & 0.0 & 7.34 & 0.0 & 0.0 \\
\bottomrule
\end{tabular}
\caption{ Results generated by different models.}
\label{tab:case_ato_metic_limit}
\end{table*}

We consider the answer without negative scores as the ground truth and employ four automatical metrics (BLEU, Meteor, Rouge-L, and Cider) to assess the quality of the generated answer. The results are shown in~\Tab~\ref{tab:so-qa-auto}. We can see that both \our and \alpaca outperform \llama across all metrics, indicating that instruction fine-tuning would help models understand human intent and generate human-like responses.
However, despite improvements over \llama, \alpaca and \our obtain relatively low scores on the four metrics.
These automatical metrics, based on token-based similarity, might not fully reflect the quality of the generated answers. For instance, consider the example in \Tab~\ref{tab:case_ato_metic_limit},  where the question is "How to get(extract) a file extension in PHP? " and the corresponding referenced answer is "<code>pathinfo()</code>". Many models (\llama-30B, \alpaca-7B, \alpaca-30B, \our-7B, and \our-30B) correctly suggest using the \textit{pathinfo()} function to extract a file extension in PHP.  However, the answers received low or inconsistent scores in BLEU, Rouge-L, and Cider, highlighting the limitations of these metrics in evaluating answer quality.  Specifically, all the answers are scored 0 in terms of BLEU, Rouge-L, and Cider, regardless of whether they are correct or not. While the answers of \alpaca-7B and \alpaca-30B outperform \llama-30B by avoiding irrelevant sentences, the Meteor score of \llama-30B is higher than that of \alpaca-7B and \alpaca-30B. Therefore, to comprehensively study the effectiveness of our approach \our, conducting human evaluations is necessary.

\subsubsection{Human evaluation}

\begin{table*}[!h]
\scriptsize
\centering
\begin{tabular}{p{1.4cm}|p{0.6cm}|p{3cm}|p{3.4cm}|p{5cm}}
\hline
\textbf{Category} & \textbf{Score} & \textbf{Scoring Criteria} &  \textbf{Example}&  \textbf{Explanation}\\
\hline
\multirow{4}{*}{\textbf{Alignment}} &
0 & The answer is entirely irrelevant, containing content that is unrelated to the question's topic. & Cats are great pets because they are low-maintenance and independent. & The answer is entirely irrelevant because it discusses pets, which have no connection to the topic of extracting file extensions in PHP.\\
\cline{2-5}
& 1 &The answer is somewhat related to the topic, but its connection to the question is weak and not directly focused on the problem. & You can determine a file type by looking at the file name. & The answer is somewhat related to the topic as it mentions file type determination, but it doesn't provide a direct solution for extracting a file extension in PHP.\\
\cline{2-5}
& 2 & The answer is relevant, displaying an understanding of the question's topic, but it may not encompass all aspects or nuances of the problem. & In PHP, you can find the file extension and name by. & The answer is relevant because it mentions the file extension, but it lacks practical solutions related to "How to".\\
\cline{2-5}
& 3 & The answer is highly relevant, demonstrating a deep comprehension of the question's topic and closely connecting to all aspects of the problem. &To find a file extension in PHP, you can split the file name with a delimiter and retrieve the last part. &  The answer is highly relevant because it suggests a method for finding file extensions in PHP, although it might not be entirely accurate. \\
\hline
\multirow{4}{*}{\textbf{Accuracy}} &
0 & The answer is entirely incorrect, providing false information or suggesting an invalid solution. &Use the `strlen()' function to find the file extension in PHP &The answer is entirely incorrect because the `strlen()' function is used to find the length of a string, not to extract a file extension..\\
\cline{2-5}
& 1 & The answer contains some correct information but also has significant inaccuracies or misconceptions. &Use the pathinfo() function. It returns the extension directly. &  The answer is partially correct, as it suggests using `pathinfo()', but it returns an array rather than the extension.\\
\cline{2-5}
& 2 & The answer is mostly accurate, with only minor errors or omissions. & Use pathinfo() in PHP to get file information, including the extension and filedir. &The answer is mostly accurate as it mentions the correct function to get file information.  However, it should be `dirname' instead of `filedir'. \\
\cline{2-5}
& 3 & The answer is completely accurate, providing correct information and a valid solution. &Use the pathinfo() function in PHP to extract the file extension: \$extension = pathinfo( \$filename, PATHINFO\_EXTENSION ); &The answer is completely accurate, providing a correct PHP function along with an example. \\
\hline
\multirow{4}{*}{\textbf{Readability}} & 
0 & The answer is extremely difficult to understand, with poor grammar, structure, or excessive jargon. & PHP file get extension method apply for find out. &The answer is extremely difficult to understand due to poor grammar and sentence structure. \\
\cline{2-5}
& 1 & The answer is somewhat difficult to understand or has some grammatical errors and unclear explanations.& php use pathinfo get file info eg extenion,basenamee,filenme &The answer is somewhat difficult to understand due to the lack of a concrete example and proper grammar.\\
\cline{2-5}
& 2 & The answer is clear, well-structured, and has only minor grammatical errors or room for improvement. & =Use the pathinfo() to extract extension: \$extension = pathinfo(\$filename, PATHINFO\_EXTENSION);& The answer provides a code example, but the readability is reduced due to the unnecessary symbol "==".\\
\cline{2-5}
& 3 & The answer is very clear, well-structured, and free from grammatical errors, making it easy to understand. & Use the pathinfo() function in PHP to extract the file extension: \$extension = pathinfo(\$filename, PATHINFO\_EXTENSION) & The answer is very clear, well-structured, and free from grammatical errors, making easy understanding.\\
\hline
\multirow{4}{*}{\textbf{Confidence}} &
0 & The rater is not at all confident in his evaluation of the answer and feels unsure about the assigned scores. & /  &  / \\
\cline{2-5}
& 1 & The rater has low confidence in their evaluation and may have doubts about the assigned scores.  & /  &  /\\
\cline{2-5}
& 2 & The rater is fairly confident in their evaluation, with only minor uncertainties about the assigned scores. & /  &  / \\
\cline{2-5}
& 3 & The rater is highly confident in their evaluation and feels certain about the assigned scores. & /  &  / \\
\hline
\end{tabular}
\caption{Scoring criteria. Examples on "How to get(extract) a file extension in PHP?".}
\label{tab:huam-score-criteria}
\end{table*}

Inspired by previous work~\cite{shi2022evaluation,shi2022race,shi2021cast}, we conduct a human evaluation to evaluate the effectiveness of \our. We randomly select 50 questions from the testing sets and collect the answers generated by the nine approaches listed in \Tab~\ref{tab:so-qa-auto}. Subsequently, we obtain 450 \texttt{<question title, question body, answer>} pairs for scoring.

Specifically, we invite 10 volunteers with excellent English abilities and over three years of software development experience. Each volunteer is asked to assign scores from 0 to 4 (with higher scores indicating better quality) to the generated answers based on four aspects: \textbf{Alignment} (the degree of understanding questions and providing relevant answers), \textbf{Accuracy} (the extent of providing correct information and valid solutions), \textbf{Readability} (grammatical correctness, the level of fluency and formatting), and \textbf{Confidence} (the degree of confidence in their evaluation). Each pair is evaluated by two volunteers, and the final score (excluding confidence) is the average of their assessments. Detailed scoring criteria, examples, and corresponding explanations are provided in \Tab~\ref{tab:huam-score-criteria}.

To ensure the reliability of the human scores, we calculated Krippendorff's alpha~\cite{hayes2007answering} and Kendall rank correlation coefficient (Kendall's Tau)~\cite{kendall1945treatment} values to assess the agreement between volunteers. 
Krippendorff's alpha value is about 0.9, and the pairwise Kendall's Tau values range from 0.75 to 0.96, indicating a high degree of agreement among the ten volunteers. Moreover, to further enhance the reliability, we had another senior volunteer double-check the labeled results with low confidence scores (less than 2).
The results of the human evaluations are shown in \Tab~\ref{tab:human_eval_results}. We can see that \llama struggles to understand the questions and provide the correct solutions. The generated answers are generally challenging to comprehend, contain grammatical errors, and lack clarity in explanations. In contrast, both \our and \alpaca outperform \llama significantly in terms of understanding questions and generating correct answers. The answers from \our and \alpaca are generally clear, well-structured, and free from grammatical errors, making them easy to understand. Remarkably, our model (\our) performs the best among all approaches in all three metrics, indicating its exceptional ability to understand questions and provide relevant and accurate answers while ensuring grammatical correctness, fluency, and good formatting.

\begin{table}[t]
\centering
\small
\setlength{\tabcolsep}{4pt}

\begin{tabular}{lccc}
\hline
\textbf{Model} & \textbf{Alignment} & \textbf{Accuracy} & \textbf{Readability} \\
\hline
LLaMa-7B & 0.11 (±0.34) & 0.02 (±0.14) & 0.08 (±0.27) \\
LLaMa-13B & 0.20 (±0.53) & 0.14 (±0.40) & 0.35 (±0.61) \\
LLaMa-30B & 0.95 (±1.13) & 0.70 (±1.04) & 1.08 (±1.21) \\
Alpaca-7B & 1.97 (±0.85) & 1.36 (±1.03) & 2.60 (±0.63) \\
Alpaca-13B & \textbf{2.52 (±0.71)} & 2.10 (±1.10) & 2.86 (±0.40) \\
Alpaca-30B & \textbf{2.52 (±0.67)} & 2.04 (±1.02) & \textbf{2.90 (±0.30)} \\
\hline
SoTaNa-7B & 2.20 (±0.69) & 1.62 (±1.09) & 2.69 (±0.48) \\
SoTaNa-13B & 2.42 (±0.80) & 2.02 (±1.10) & 2.71 (±0.59) \\
SoTaNa-30B & \textbf{2.52 (±0.74)} & \textbf{2.16 (±0.92)} & \textbf{2.90 (±0.30)} \\
\hline
\end{tabular}
\caption{Human evaluation results.}
\label{tab:human_eval_results}
\end{table}

\subsection{Experiment on Code Summarization and Generation}

\subsubsection{Overall results}
\begin{table}[ht]
\small
    \centering
    \setlength{\tabcolsep}{3pt}
   
    \begin{tabular}{l c cccc}
    \toprule
      \multirow{1}{*}{Model}  &  \multicolumn{1}{c}{Code Generation}  
      &\multicolumn{4}{c}{Code Summarization} \\
    \cmidrule(r){2-2} \cmidrule(r){3-6}
    &P@1  &BLEU &MET. &Rou. &Cid. \\ 
\midrule
\llama-7B & 10.5 &0.29 &2.41 &2.24 &0.00 \\ 
\llama-13B &15.8 &0.33 &3.17 &3.44 &0.01 \\  
\llama-30B &21.7 &0.89 &5.21 &6.34 &0.01 \\ 
\alpaca-7B &10.37&3.80 &12.97 &19.71 &0.31 \\  
\alpaca-13B &12.20&3.67 &12.67 &19.88 &0.29 \\ 
\alpaca-30B &18.90&\textbf{4.69} &14.51 &22.25 &\textbf{0.48} \\  
\hline
\our-7B & 10.97 & 3.46 &14.32 &19.96 &0.23 \\
\our-13B & 18.30 &3.71 &13.02 &19.52 &0.27 \\ 
\our-30B &\textbf{23.17} &\textbf{4.69} &\textbf{15.29} &\textbf{22.93} &0.47  \\
\bottomrule
\end{tabular}
 \caption{ Results on code summarization and generation.}
\label{tab:code-sum-gen}
\end{table}

To evaluate the effectiveness of our model in understanding and generating code, we conducted experiments on two benchmarks and compared our model \our with those of \llama and \alpaca. The experimental results are shown in \Tab~\ref{tab:code-sum-gen}. We can see that larger model sizes generally lead to better performance on both code summarization and generation. Compared to \llama, \alpaca and \our show significant improvements in code summarization. This suggests that fine-tuning models with human instructions can enhance their ability to understand and generate natural language sentences resembling human-like descriptions. Moreover, \our demonstrates an improvement in \llama's code generation capability, whereas \alpaca's fine-tuning appears to decrease \llama's code generation ability. This indicates that fine-tuning models using general-purpose data could potentially impair their code generation capabilities. On the other hand, fine-tuning with software engineering-related data, as done in our approach, enhances the model's code generation proficiency. In summary, our experiments demonstrate that our model benefits from instruction-based tuning, leading to improvements in both code summarization and generation tasks. Additionally, fine-tuning software engineering domain-specific data proves to be advantageous for code generation.

\subsubsection{The impact of data volumes}

\begin{figure*}[ht]
    \centering
    
\begin{minipage}[t]{0.43\linewidth}
        \centering
        \includegraphics[width=1\linewidth]{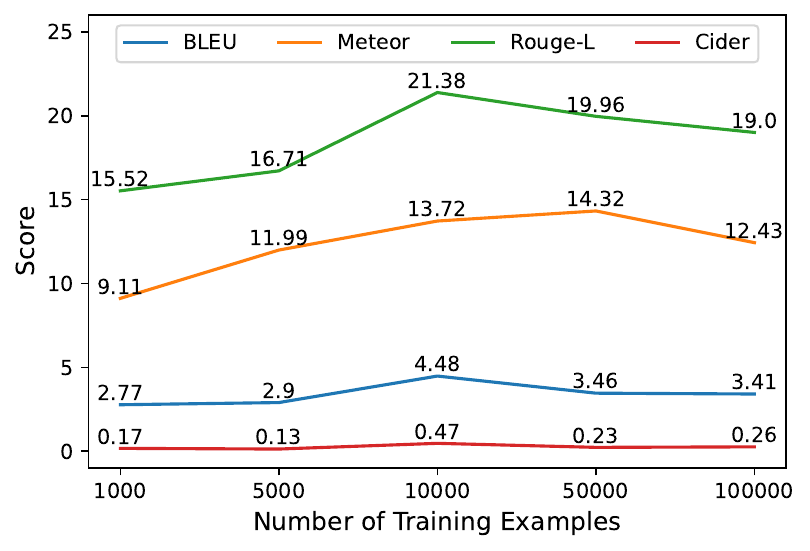}
        \caption{\our{}-7B on code summarization}
    \end{minipage}
    \begin{minipage}[t]{0.43\linewidth}
        \centering
        \includegraphics[width=1\linewidth]{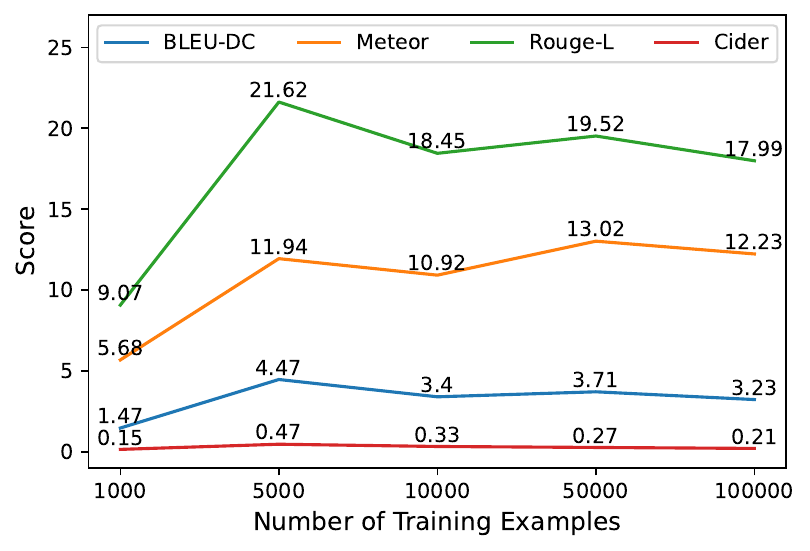}
       \caption{\our{}-13B on code summarization}
    \end{minipage}
    
    \begin{minipage}[t]{0.43\linewidth}
        \centering
        \includegraphics[width=1\linewidth]{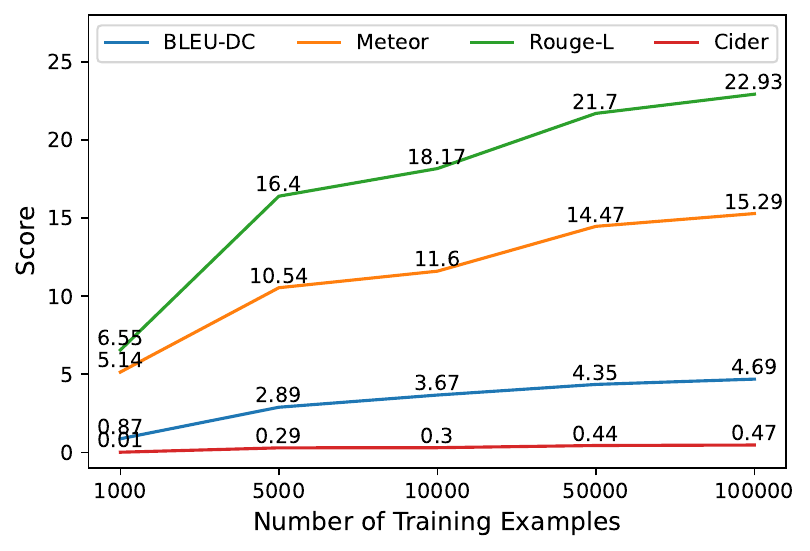}
         \caption{\our{}-30B on code summarization}
    \end{minipage}
        \begin{minipage}[t]{0.43\linewidth}
        \centering
        \includegraphics[width=1\linewidth]{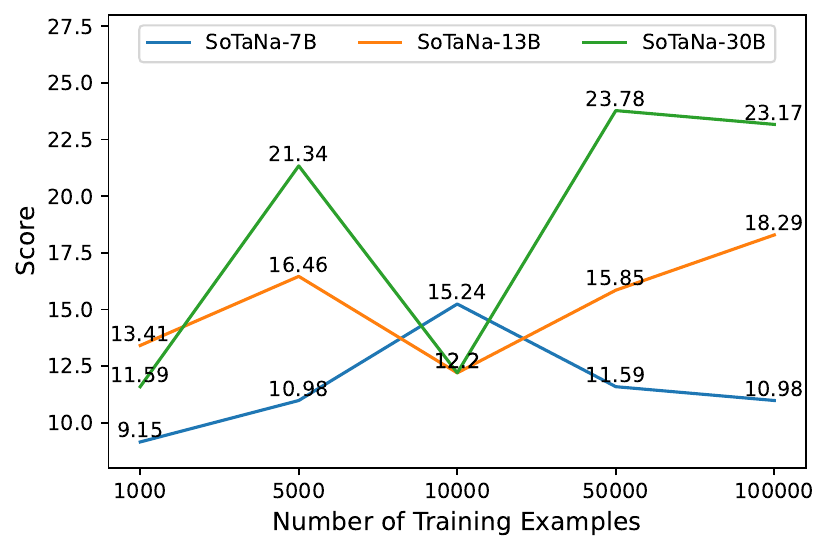}
         \caption{\our{} on code generation}
    \end{minipage}

     \label{fig:diff_r}   
\caption{The impact of different data size.}
\label{fig:diff_params}
\end{figure*}

We conduct further investigation into the impact of varying the volume of generated data on model performance. Specifically, we tune the \llama model using datasets of different sizes: 1k, 5k, 10k, 50k, and 100k generated examples. Subsequently, we evaluate the models on both code summarization and code generation tasks, and the results are shown in \Fig~\ref{fig:diff_params}.

Interestingly, we see that the performance does not consistently increase with the increase in data size, except for \our-30B on code summarization, which shows improvement. One possible reason for this inconsistency could be the issue with the evaluation metrics. As we discuss in \Sec~\ref{sec:auto-metic-qa}, the automatic metrics might not effectively measure the quality of the generated results.

Additionally, we notice that the impact of varying data size on model performance is not consistent across different model sizes. That is, conclusions or findings drawn for one model size cannot be directly applied to another size. For instance, \our-13B achieves the best performance on code summarization when using 5K data, while \our-7B and \our-30B did not exhibit the same trend. For code generation, \our-7B performs exceptionally well when trained on 10K data, whereas \our-7B and \our-30B show the worst performance with the same dataset size.

The results indicate the importance of careful consideration when selecting the data size and model configuration for specific tasks. It also emphasizes the necessity of using appropriate evaluation metrics to accurately assess the model's performance on some code-related tasks.

\section{Discussion}

The work most closely related to our research is the StarChat~\cite{starchat} project. They fine-tune a model called StarCode, which is designed specifically for code, using general-purpose data to make StarCoder~\cite{li2023starcoder} capable of handling dialogue. In contrast, our approach centers around using software engineering-related data to fine-tune a general-purpose large language model, with the aim of creating a software development assistant.

\section{Threats to Validity}

\textbf{Data Quality.} Another potential concern lies in the data generation process using ChatGPT. While we have filtered out some low-quality datasets, such as instructions with less than 3 words, we acknowledge that human checks for the correctness of the generated data were not performed. To improve the quality of generated data, future work can incorporate human checks or propose alternative approaches to ensure data accuracy and reliability.

\textbf{Evaluation Datasets.} The experiments have been conducted on widely-used datasets; however, there are other datasets available for each downstream task. These alternative datasets may differ in construction methods and corpus sizes, potentially leading to variations in model performance. To enhance the robustness of our findings and conclusions, further experiments on a broader range of datasets can be carried out to validate the generalizability of the results.

\textbf{Evaluation Metrics.} We have utilized commonly-used metrics to assess the performance of the models. However, it is crucial to recognize that these metrics may have inherent limitations. For example, metrics like BLEU and METEOR rely on textual similarity and may not fully capture the semantic similarity between two sentences.
To address these limitations and obtain a more comprehensive evaluation, we also conducted human evaluations. However, it's worth noting that human evaluations can be labor-intensive and time-consuming. In future research,  we will explore new automatic evaluation metrics that are more aligned with human perception.

\section{Conclusion}

This paper presents \our, an open-source software development assistant designed to meet the increasing demand for effective software development tools. By leveraging Large Language Models (LLMs), \our generates high-quality instruction-based data tailored for software engineering tasks. It employs a parameter-efficient fine-tuning approach to enhance the capabilities of the \llama open-source foundation model. Through comprehensive evaluation, including human evaluation, \our demonstrates its efficacy in assisting developers by providing accurate answers to diverse Stack Overflow queries. It outperforms existing language models in understanding human intent and generating contextually appropriate responses specific to software engineering tasks.

In future work, we aim to introduce a benchmark to further evaluate LLMs as open-source software development assistants. This benchmark will provide a standardized and systematic approach for assessing the performance of various language models in the context of software engineering, enabling better comparison and progress tracking.

\bibliographystyle{acl_natbib}
\bibliography{ref}

\end{document}